\documentstyle[aps,prl,graphics,times]{revtex}
\newcommand{\micron}{$\mu\text{m}$}
\begin{document}
\draft

\author{{D. A. Reis}$^{1,2}$, {M. F. DeCamp}$^{1}$, {P. H. 
Bucksbaum}$^{1,2}$, {R. Clarke}$^{1}$, {E. Dufresne}$^{1}$,
{M. Hertlein}$^{1}$,  
{R. Merlin}$^{1,2}$,  {R. Falcone}$^{3}$, {H. 
Kapteyn}$^{4}$, {M. Murnane}$^{4}$, {J. Larsson}$^{5}$, {Th. 
Missalla}$^{5}$, {J. Wark}$^{6}$ }
\address{$^{1}${Department of Physics, University of Michigan, Ann Arbor, 
MI 48109-1120}\\
$^{2}${Center for Ultrafast Optical 
Science, University of Michigan, Ann Arbor, MI 48109-2099}\\
$^{3}${Department of Physics University of California, Berkeley, 
CA 94720-7300}\\
$^{4}${JILA  and Department of Physics, University of Colorado 
and National Institute of Standards and Technology, Campus Box 440, 
 Boulder, CO 80309-0440}\\
$^{5}${Atomic Physics Division, Lund Institute of Technology, 
P.\ O.\ Box 118, SE-221 00,
Lund, Sweden}\\
$^{6}${Department of Physics, Clarendon Laboratory, University 
of Oxford, Parks Road, Oxford, OX1 3PU, U.K.}}
\date{\today}
\title{Probing impulsive strain propagation with x-ray pulses}

\maketitle
\begin{abstract}
  Pump-probe time-resolved x-ray diffraction of allowed and nearly
  forbidden reflections in InSb is used to follow the propagation of a
  coherent acoustic pulse generated by ultrafast laser-excitation. 
  The surface and bulk components of the strain could be
  simultaneously measured due to the large x-ray penetration depth. 
  Comparison of the experimental data with dynamical diffraction
  simulations suggests that the conventional model for impulsively
  generated strain underestimates the partitioning of energy into
  coherent modes.
\end{abstract}


The absorption of ultrafast laser pulses in opaque materials
generates coherent stress when the pulse length is short compared with
time for sound to propagate across an optical penetration depth
\cite{ultraShortAcousticRev}.  The resulting strain field consists of
both a surface component, static on time scales where thermal
diffusion can be ignored, and a bulk component that propagates at the
speed of sound (coherent acoustic phonons).  This strain is typically
probed by optical methods that are sensitive primarily to the phonon
component within the penetration depth of the
light~\cite{ultraShortAcousticRev,AcousticMeasRev}.  However, such
methods give little information about the surface component of the
strain, and, moreover, they are unable to give a quantitative measure
of the strain amplitude.

Due to their short wavelengths, long penetration depths, and
significant interaction with core electrons, x-rays are a sensitive
probe of strain.  We note that coherent lattice motion adds sidebands
to ordinary Bragg reflection peaks due to x-ray Brillouin scattering
if the momentum transfer is large compared to the Darwin width,
equivalent to phonons of GHz frequency for strong reflections from
perfect crystals.  This effect was demonstrated many years ago with
acoustoelectrically amplified phonons using a conventional x-ray
tube~\cite{colellaBragg}.

With the recent availability of high brightness short-pulse hard x-ray
sources, including third generation synchrotron sources and optical
laser based sources
\cite{laserPlasmaSource,thomsonSource,PonderomotiveSlice}, coherent
strain generation and propagation can now be probed by x-ray methods
in both the frequency and time domains.  Recently, time-resolved
diffraction patterns of cw ultrasonically excited crystals were
obtained with a synchrotron source\cite{esrfMHz}.  Other experiments
have employed picosecond time-resolved x-ray diffraction to study
transient lattice dynamics in metals~\cite{Gold}, organic
films~\cite{fsOrganic}, and impulsive strain
generation and melting in semiconductors%
~\cite{GaAsNature,GeScience,berkeleyThomsonPRL,berkeleyMeltPRL,GePRL}. 
In particular Rose-Petruck \emph{et al.}\ \cite{GaAsNature} demonstrated
transient ultrafast strain propagation in GaAs by laser-pump
x-ray-probe diffraction.  In that experiment, x-rays were diffracted
far outside the Bragg peak; however, no oscillations in the
diffraction efficiency were detected, and the data were consistent
with a unipolar strain pulse.  In a similar experiment, Lindenberg et
al.\ \cite{berkeleyMeltPRL} detected oscillations in the sidebands for
an asymmetrically cut InSb crystal using a streak camera.  These
oscillations were due to lattice compression and were probed for
discrete phonon frequencies in the near surface regime.

In this Letter, we report on the modulation of x-ray diffraction from
(111) InSb due to impulsively generated coherent acoustic phonons.  We
combined ultrafast pump-probe techniques with high resolution x-ray
diffraction to build up a comprehensive time and frequency resolved
picture of the induced strain.  Because the x-rays probed deeper into
the material than the laser penetration depth, the bulk, phonon
component of the strain could be distinguished from the surface
component as it propagated into the crystal.  The data reveal both the
compression and rarefaction of the associated strain wave.  When
compared with the standard model of strain
production\cite{thomsenStrain}, our data suggest that the bulk
component of the strain is twice the amplitude predicted by this
model.

Both the allowed (strong) and nearly forbidden (weak) reflections from
the 111 and 222 crystallographic planes of the face-centered cubic
crystal were used.  We believe this to be the first study of the
effect of ultrasound on the diffraction in a nearly forbidden
reflection offering important advantages in terms of
the x-ray absorption depth and sensitivity to low frequency phonons. 

The experiments were conducted at the Michigan/Howard/Lucent
Technologies-Bell Labs collaborative access team (MHATT-CAT) 7ID
insertion device beamline at the Advanced Photon Source (APS).  The
x-rays from the undulator were monochromatized by a cryogenically
cooled silicon 111 double crystal monochromator.  The photon energy
was set to 10 keV and the bandwidth was approximately 1.4~eV. The
x-rays were \emph{p}-polarized with respect to the sample resulting in
a reduction of the Bragg reflectivity compared to the reflectivity for
\emph{s}-polarization\cite{pPolEffect}.  The scattering plane of the
sample was also perpendicular to that defined by the monochromator;
however, the effect of dispersion on the reflections is expected to be
small.  The beam was collimated with slits before both the
monochromator and the sample.

Ionization chambers monitored the x-ray flux before and after the
sample.  A silicon avalanche photodiode (APD), with a rise time of \(
< \)3 ns after preamplification~\cite{baronAPD}, provided sufficient
temporal resolution to gate a single x-ray pulse from the 150 ns pulse
spacing in the APS singlets mode.  The data were background subtracted
to remove an angle dependent DC offset in the APD due to overshoot in
the detector electronics from the previous x-ray bunch.

A commercial ultrafast 840 nm kHz Ti:sapphire laser impulsively
excited large amplitude coherent acoustic phonon fields in the sample. 
The laser was phase-locked to a single x-ray pulse in the synchrotron
with a timing jitter less than the x-ray pulse duration of 90 ps
(FWHM).  A digital phase shifter in the reference of the feedback loop
provided a variable delay with 19 ps precision.  The \emph{p}-polarized
laser pulse was focused to 1 x 1.5 mm$^{2}$ at the sample, with an
incident angle of 50 degrees, on the same side of normal as the
incident x-ray pulse.  The fluence was 10--12 \(
\text{mJ}/\text{cm}^{2} \).  The temporal structure of the laser
consisted of two 70 fs pulses separated by less than a picosecond. 
This substructure is not expected to play a role in the strain
dynamics.

\begin{figure}
     \includegraphics{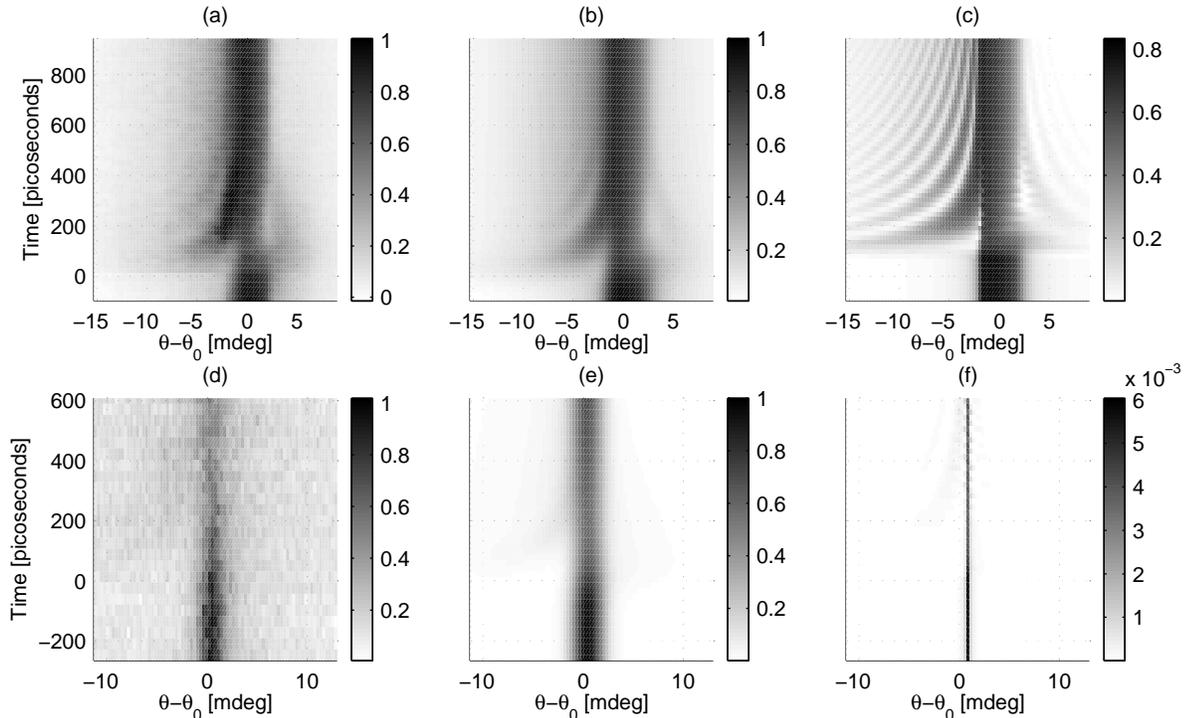}%
    \caption{Time-resolved rocking curves of impulsively strained InSb.
    (a),(d) Data and (b),(e) dynamical diffraction simulation for a
    0.51\% peak strain with a 100 nm laser absorption depth and a 100
    ps, 1.25 mdeg (2.5 mdeg for the 222) FWHM Gaussian resolution
    functions (normalized to the peak unstrained signal); and (c),(f)
    Dynamical diffraction simulation for the reflectivity of a
    monochromatic x-ray pulse with ideal temporal resolution, all for the
    111, 222 reflections respectively.}
    \label{f:results}
\end{figure}

Our results are summarized in Fig.\ \ref{f:results}.  Figure
\ref{f:results}(a) shows the measured time-resolved rocking curve for
the symmetric 111 Bragg reflection following ultrafast laser
excitation.  The vertical axis is the time delay between the x-ray and
the laser pulses.  The time when the strain is first observed is
defined as $t=0$.  The horizontal axis is the x-ray incident angle
with respect to the center of the Bragg peak, $\theta_{0}$, in the
laser-on condition.  After the arrival of the laser pulse, the rocking
curve broadens and shifts slightly towards smaller angles.  During the
first few hundred picoseconds, oscillations occur on the low angle
side of the peak while on the high angle side there is both transient
broadening and, for certain angles, decreased reflectivity.  At longer
delays only the broadening at negative angles persists, and no
oscillations are evident.

Much of the behavior in Fig.\ \ref{f:results}(a) can be understood by
kinematical diffraction.  In the kinematic approximation, x-rays
scattered at a given offset angle diffract from regions of material
with appropriate crystallographic plane spacing.  If the spacing
between the planes is changing in time, then an offset
\(\theta-\theta_{B} \) occurs due to Brillouin scattering of the
x-rays from phonons of wavevector \( q =
d^{-1}(\theta-\theta_{B})\cot\theta_{B} \) where \( \theta_{B} \) is
the Bragg angle and \( d \) is the unperturbed plane spacing
\cite{centralAngle}.  Smaller angles thus correspond to lattice
expansion and larger angles correspond to compression.

Oscillations occur in the diffracted signal because the phonons are
coherent (since the stress is impulsive).  A fast Fourier transform of
the data for the expanded material gives a dispersion relation for the
phonons consistent with the speed of sound for longitudinal acoustic
modes propagating along the (111) direction.  The temporal response of
the material is expected to extend to 40 GHz assuming a 100 nm laser
absorption depth and a 3880 m/s speed of sound; however the x-ray
pulse length, timing jitter and dephasing (due to a small momentum
spread of the x-rays) limit the detection of oscillations to
approximately 10 GHz.  Nonetheless higher frequency phonons are
revealed by the existence of diffracted x-rays well outside the main
Bragg peak corresponding to acoustic phonons up to approximately 40
GHz.

The longitudinal coherence length of the x-rays is 0.9 \micron\
allowing for interference between the x-rays diffracted from the
strained and unstrained portions of the crystal.  The absorption depth
of 10 keV x-rays in InSb is approximately 2.3 \micron\ for the
symmetric 111 reflection, corresponding to 600 ps at the acoustic
velocity.  In a strong reflection, the depletion of the input beam
(extinction) limits the penetration depth.  This extinction broadens
the width of the diffraction peak and thus sets a lower limit on the
phonon wavevector which can be resolved as a sideband.  Strain
persists on the expansion side well after the phonons traverse the
absorption depth (as further indicated by the damping of any
oscillations).  This indicates that expanded material is left in the
wake of the sound front due to surface heating from the laser.   
The slow decay of this surface strain occurs over a few hundred nanoseconds 
and is due to thermal diffusion; however, this is unimportant on the
time scale of Fig.\ \ref{f:results}(a).

In order to study strain propagation over longer times, the x-ray
must penetrate deeper into the bulk.  At a fixed x-ray energy this 
requires either a weaker reflection or larger momentum transfer 
(larger Bragg angles).
The 222 reflection for InSb is structurally
forbidden except for a small difference in the scattering factors of
the In and Sb atoms;  the reflection is weak enough that extinction
does not play a role, and the probe depth is limited by the absorption
depth, which at 10 keV corresponds to approximately
4.5 \micron .

Figure \ref{f:results}(d) shows the time-resolved diffraction
measurement of the symmetric 222 peak.  In this data the
time delay was scanned with 38~ps steps.  The collimation slits were
opened slightly wider than for the 111 data in order
to increase the incident x-ray flux, at the expense of a slight
decrease in the angular resolution.  The peak of the rocking curve
decreases in amplitude after laser excitation, with a time constant
limited by our temporal resolution.  After the initial drop, the
diffraction efficiency again begins to build up, returning to the
pre-strained value well before the arrival of the next laser pulse.  In
addition, after excitation there is a slight increase in the
diffraction efficiency on the expansion side of the rocking curve. 
The signal to noise for the 222 reflection was insufficient to resolve
any compression dynamics.

The data for both reflections are compared with simulations that model
dynamical diffraction theory for a single
reflection~\cite{zachariasen,batterman-cole} in the presence of
strain~\cite{takagi,taupin}.  The lack of inversion symmetry in InSb
is ignored, except in the determination of the structure factor for
the 222 reflection.  The rocking curves are calculated for a
particular strain profile using the method given by Wie \emph{et al.}\
\cite{WieSimulation}.  The strain is calculated using the model of
Thomsen \emph{et al.}\ \cite{thomsenStrain}, who solved the elastic
equations for an instantaneous and exponentially decaying stress due
to short-pulse laser absorption.  In our case this model predicts both
a 0.51\% strain across the 100 nm absorption depth and a 0.51\%
peak-to-peak bulk acoustic strain.  Lindenberg \emph{et al.}\
\cite{berkeleyMeltPRL} found it necessary to include both an
instantaneous stress due to the effect of free carriers on the
deformation potential and a 12 ps delayed stress which they attribute
to electron-phonon coupling.  Similar delays are found in Ref.\
\cite{GaAsNature,berkeleyThomsonPRL}.  Such delays are on the order of
the phonon propagation time across the laser penetration depth and are
unresolved due to the temporal length of the x-ray probe.

Figures \ref{f:results}(c) and \ref{f:results}(f) show results of the
simulations for the predicted 0.51\% strain.  To account for
experimental resolutions, the calculations shown in Fig.\
\ref{f:results}(b) and Fig.\ \ref{f:results}(e) were convolved with
Gaussians of 1.25 mdeg for 111, and 2.5 mdeg for 222, and 100 ps FWHM.
In Fig.\ \ref{f:results}(b), a small incoherent 14 mdeg FWHM Gaussian
pedestal was added to better reproduce the data (15\% the amplitude of
the peak).
%

\begin{figure}
     \includegraphics{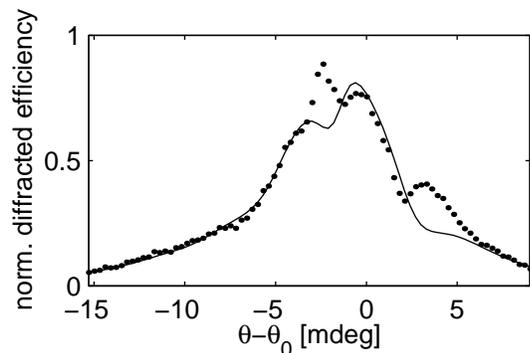}
    \caption{Rocking curve of the 111 peak at  150 ps time delay 
    (normalized to the peak unstrained value). 
    The solid circles are the data and the line is the simulation.  
    The discrepancy is interpreted as due to an excess of energy in 
     coherent acoustic modes.}
    \label{f:lineout}
\end{figure}

While the simulations reproduce many of the features in the data, they
underestimate the phonon contribution of the strain.  This is
particularly evident in the rocking curve of Fig.\ \ref{f:lineout}. 
To model the strain field more accurately, we allowed for separate
partitioning of energy into surface and bulk components.  A least
squares fit of the $\sim$6000 data points for Fig.\ \ref{f:results}(a)
yielded a 0.39\% amplitude surface strain with a 0.76\% peak-to-peak
acoustic strain with a $\chi^{2}_{dof}=0.5$.  By comparison, the model of
Thomsen \emph{et al.} gives $\chi^{2}_{dof}=0.8$.  The relative
statistical error in the strain determination is at the percent level,
indicating that our data is limited by systematic errors.

We believe the data presented here to be the first x-ray diffraction
probe of a bipolar ultrasonic pulse during its propagation into the
bulk.  We note that the time over which the pulse can be followed here
was limited by absorption and extinction of the x-ray probe, and that
in the Laue geometry the probe depth may be extended to the entire
bulk of the crystal.  Data collected with such a technique are
currently under analysis for the case of a thick germanium single
crystal.  We further note that while the detection of the direct
temporal response of the material was limited to 100 ps, shorter times
could be probed due to the momentum sensitivity of x-ray diffraction. 
Improvements in the direct temporal resolution will allow for a more
detailed study of the stress generation and the partitioning of energy
into the lattice.  Finally, extension of the techniques, discussed
here, to the terahertz regime using coherent optical phonons may be
used to slice sub-picosecond x-ray pulses from the somewhat longer
pulse structure of third generation x-ray sources~\cite{BraggSwitch}.

\begin{acknowledgments}
We would like to thank A. Baron for providing us with the preamplifier
circuit for the APD and E. Williams, S. Dierker and W. Lowe for their
experimental support.  This work was conducted at the MHATT-CAT
insertion device beamline on the Advanced Photon Source and was
supported in part by the U.S. Department of Energy, grants
DE-FG02-99ER45743 and DE-FG02-00ER15031.  Use of the Advanced Photon
Source was supported by the U.S. Department of Energy, Basic Energy
Sciences, Office of Energy Research, under Contract No. 
W-31-109-Eng-38.  One of us (DR) acknowledges support from the
National Science Foundation CUOS Fellows program.
\end{acknowledgments}
\bibliographystyle{prsty}
\bibliography{timeResolvedXray}

\begin{thebibliography}{10}
\bibitem{pPolEffect}
The peak reflectivity for \emph{p}-polarization is \protect$\cos^2
  2\theta_B\protect$ that of the corresponding reflectivity for
  \emph{s}-polarization. This amounts to a 11\% (38\%) reduction in the
  reflectivity for the 111 (222) reflection respectively.
\bibitem{centralAngle}
In the Bragg geometry there is a slight difference between the center of the
  Bragg peak, \protect$\theta_0\protect$, and the Bragg angle, defined through
  \protect$\sin\theta_B = \lambda/2d\protect$. This is due to the small
  difference in the index of refraction from unity.
\end{thebibliography}

\end{document}